\documentclass[10pt, conference]{IEEEtran}
\IEEEoverridecommandlockouts
\usepackage[utf8]{inputenc}
\usepackage{amsmath,amssymb,amsfonts}
\usepackage{textcomp}
\usepackage[table,xcdraw]{xcolor}
\def\BibTeX{{\rm B\kern-.05em{\sc i\kern-.025em b}\kern-.08em
    T\kern-.1667em\lower.7ex\hbox{E}\kern-.125emX}}

\usepackage{algorithm}
\usepackage[noend]{algpseudocode}
\usepackage{float}
\usepackage{caption}
\usepackage{multirow}  
\usepackage{amsmath}
\usepackage{titlesec}
\titlespacing*{\section}{0pt}{1.1\baselineskip}{\baselineskip}
\usepackage{cellspace, hhline}
\usepackage{mathtools}
\usepackage{subfig}

\usepackage{breqn}
\usepackage[colorlinks=true, citecolor=blue,linkcolor=blue,bookmarks=false]{hyperref}
\usepackage{amssymb}
\usepackage[export]{adjustbox}
\usepackage[symbol]{footmisc}
\def\algbackskip{\hskip-\ALG@thistlm}

\usepackage[switch]{lineno}
\newcommand\nnfootnote[1]{%
  \begin{NoHyper}
  \renewcommand\thefootnote{}\footnote{#1}%
  \addtocounter{footnote}{-1}%
  \end{NoHyper}
}

\newcommand\githubref[1]{%
  \renewcommand\thefootnote{}\footnote{#1}%
  \addtocounter{footnote}{-1}%
}

\makeatletter 
\newcommand{\algmargin}{\the\ALG@thistlm}
\makeatother
\algnewcommand{\parState}[1]{\State%
    \parbox[t]{\dimexpr\linewidth-\algmargin}{\strut\hangindent=\algorithmicindent \hangafter=1 #1\strut}}

\usepackage{xcolor,soul,lipsum}
\newcommand{\myul}[2][black]{\setulcolor{#1}\ul{#2}\setulcolor{black}}

\DeclareRobustCommand*{\IEEEauthorrefmark}[1]{%
\raisebox{0pt}[0pt][0pt]{\textsuperscript{\footnotesize #1}}%
}
\usepackage[nodisplayskipstretch]{setspace}
\usepackage[style=ieee]{biblatex} 
\usepackage{caption}
\usepackage{multicol}
\usepackage{adjustbox}
\bibliography{references}
\begin{document}

\title{Functional Protein Structure Annotation Using a Deep Convolutional Generative Adversarial Network}

\author{\IEEEauthorblockN{Ethan Moyer\IEEEauthorrefmark{1}\IEEEauthorrefmark{*}, Jeff Winchell\IEEEauthorrefmark{2,}\IEEEauthorrefmark{4}, Isamu Isozaki\IEEEauthorrefmark{2}, Yigit Alparslan\IEEEauthorrefmark{2}, Mali Halac\IEEEauthorrefmark{3}, Edward Kim\IEEEauthorrefmark{2}}

\IEEEauthorblockA{
\IEEEauthorrefmark{1}School of Biomedical Engineering, Drexel University, PA\\
\IEEEauthorrefmark{2}College of Computing \& Informatics, Drexel University, PA\\
\IEEEauthorrefmark{3}College of Engineering, Drexel University, PA\\
\IEEEauthorrefmark{4}College of Arts \& Sciences, Drexel University, PA\\
Email: \{ ejm374, jmw479, imi25, ya332, mh3636,  ek826 \}@drexel.edu}}

\maketitle
\begingroup\renewcommand\thefootnote{\textsection}
\endgroup

\begin{abstract}
Identifying novel functional protein structures is at the heart of molecular engineering and molecular biology, requiring an often computationally exhaustive search. We introduce the use of a Deep Convolutional Generative Adversarial Network (DCGAN) to classify protein structures based on their functionality by encoding each sample in a grid object structure using three features in each object: the generic atom type, the position atom type, and its occupancy relative to a given atom. We train DCGAN on 3-dimensional (3D) decoy and native protein structures in order to generate and discriminate 3D protein structures. At the end of our training, loss converges to a local minimum and our DCGAN can annotate functional proteins robustly against adversarial protein samples. In the future we hope to extend the novel structures we found from the generator in our DCGAN with more samples to explore more granular functionality with varying functions. We hope that our effort will advance the field of protein structure prediction.


\end{abstract}

\nnfootnote{\IEEEauthorrefmark{*} Corresponding author}
\githubref{All source code is open-sourced at \href{https://github.com/drexelai/protein-nets}{\color{blue} \myul[blue] {GitHub.}}}

\section{Introduction} \label{Introduction}

One of the goals of metagenomics is to identify the functions of proteins present in a given sample. Two commonly used methods to determine the protein functions are 1. to compare the amino acid sequence of a protein with the functionally annotated sequences present in protein sequence databases, 2. to compare the 3-D structure of a protein against those of the protein structure databases  \cite{lietal}. Thanks to the recent advances in computational tools and techniques especially applications of machine learning in the field of metagenomics, there is a growing number of annotations of proteins available. 

The inverse problem, determining the 3-D structure of a protein for a given function, is a young field which has attracted the interest of researchers as engineering of proteins with certain functions has promising applications in biotechnology and medicine \cite{protstructpredict}. Design of such proteins may lead to novel therapeutic agents such as custom designed signaling proteins that will allow us to give specific instructions to cells \cite{Gurevich2014}.

To address this issue, we propose an implementation of a Deep Convolutional Generative Adversarial Network using a protein data set obtained from Protein Data Bank (PDB) database. 

This paper is organized such that \autoref{relatedwork} discusses related work, \autoref{methodology} discusses the implemented search algorithms, \autoref{results} reports the experiments and results, \autoref{conclusion} concludes the paper by going over the important findings and discusses future work.

\section{Related Work} \label{relatedwork}
Functionality of a protein and its structure are tightly coupled. Understanding the 3-D structure of a protein can give us knowledge regarding its functionality.

\subsection{Protein Structure Prediction}
In the literature,  we see that X-ray crystallography and Nuclear Magnetic Resonance (NMR) are used to determine the 3-D structure of a protein \cite{ilari2008protein}. By emitting X-ray onto protein and measuring the diffractions and scatters, one can measure the density of molecules in a protein.  
NMR technique is faster compared to X-ray crystallography, but it is only used on proteins that have less than 150 amino acids \cite{goodman2000relationships}. Therefore, developing computational models that would predict protein structure is a crucial need \cite{moyer2020machine}. 

For this reason, identifying the native three-dimensional (3-D) structure of a protein is a common problem in bioinformatics and has applications in drug design, protein engineering, and protein annotation. Previous methods of 3-D structure prediction have focused on energy minimization to find thermodynamically favored structures and the results are assessed in comparison to the free energy of the native structure \cite{dill2008protein}. 

A few works in energy prediction have successfully used Convolution Neural Networks (CNN) to predict the energy between each of the bonds in a structure \cite{yao2017intrinsic}. Moreover, some have attempted to quantify the relative energy deviation of a decoy structure from its native, or most optimally folded, structure \cite{moyer2020measuring}. This latter method is displayed in \autoref{fig:methods}a where the red to blue gradient corresponds to a measurement of energy deviation from a decoy structure to a native structure.

\subsection{Functional Protein Annotation}
The function of a protein is closely tied to its structure. A similar sequence of amino acids between two proteins can imply an identical or similar function. However there are cases of even a single amino acid change entirely changing the function of a protein \cite{schaefer2012predict}. As such, additional criteria beyond protein structure is needed to predict the function of a protein. To simplify this task, there is a large body of work that focuses on identifying "structural motifs," or certain protein structures and amino acid sequences which are found in many proteins with a specific function. It should be noted that the presence of a structural motif in that protein does not necessarily indicate that protein has a certain function.

A more general question is whether a protein is functional or non-functional. Since functionality is more or less indicative of native folding, one would suspect that a search of functional proteins to computationally expensive. To put it in perspective, the protein structure search space of an $n$-lengthed amino acid sequence has $4^n$ permutations. Each individual amino acid sequence may have a range of unique structures in which the protein can function relatively well and a select few in which it functions most optimally. An exhaustive search is therefore unrealistic and effort should be put into recognizing relationships between functionally annotated structures that are already identified using X-ray crystallography and NMR.

\subsection{Generative Adversarial Network}
Generative Adversarial Networks (GANs) are first introduced by Ian Goodfellow and have seen wide adaptations \cite{goodfellow2014generative}. Even though noise cancellation was the first purpose of GANs, the field expanded onto developing conditional GANs and has seen wide adaptations in style transfer \cite{isola2018imagetoimage}, image generation \cite{zhu2020unpaired}, audio generation   \cite{alparslanspeech2020} \cite{sparsitypaper}.\\
 
A Generative Adversarial Network (GAN) can be thought of as a zero-sum game between two networks: 1) One to discriminate between real and fake data samples and 2) one to generate data samples that fool this discriminator. This dynamic is illistrated in \autoref{fig:methods}d. In mathematical terms, this corresponds to the minimax game:
\begin{align*}
     \min\limits_G \max\limits_D V(D,G) = &\mathbb{E}_{x\sim p_{\text{data}}(x)}[\log D(x)]\\&+\mathbb{E}_{z\sim p_z(z)}[\log(1-D(G(z))]
\end{align*}
where $x$ is a vector of real data samples, $z$ is a latent representation of a fake data sample, $G$ is the generative network, and $D$ is the discriminative network\cite{goodfellow2014generative}. Often, the generator uses random noise in order to create seemingly novel samples that are increasingly indistinguishable real samples. In training both of these models simultaneously, they are able to both become more accurate with discrimination and generation. 

\subsection{Deep Convolutional GAN (DCGAN)}

DCGANs merge the areas of convolutional neural networks and the GAN architecture described in C. It extends the standard GAN architecture by replacing the fully-connected generator and discriminator networks with deep convolutional neural networks. 

Convolutional Neural Networks (CNNs) are useful for classifying images, especially over their non-convolutional counterparts because the convolution operation preserves spatial properties of images by working with 2D representations. In contrast, non-convolutional networks require 1D representations and the image is "flattened" (the rows/columns of the image are concatenated). \cite{DBLP:journals/corr/abs-1905-03288} and \cite{SHARMA2018377} provide more in-depth discussions of how CNNs are used for image classification/recognition.

The use and deployment of a GAN or DCGAN is highly dependent on the problem at hand. In one case, this model could be used in order to discriminate between otherwise indistinguishable samples, such as \cite{}. In another case, a model may be deployed for the generation of unique samples. Our work focuses on the latter.

\section{Methodology} \label{methodology}

\begin{figure*}[t]
    \centering
    \includegraphics[width=0.6\textwidth]{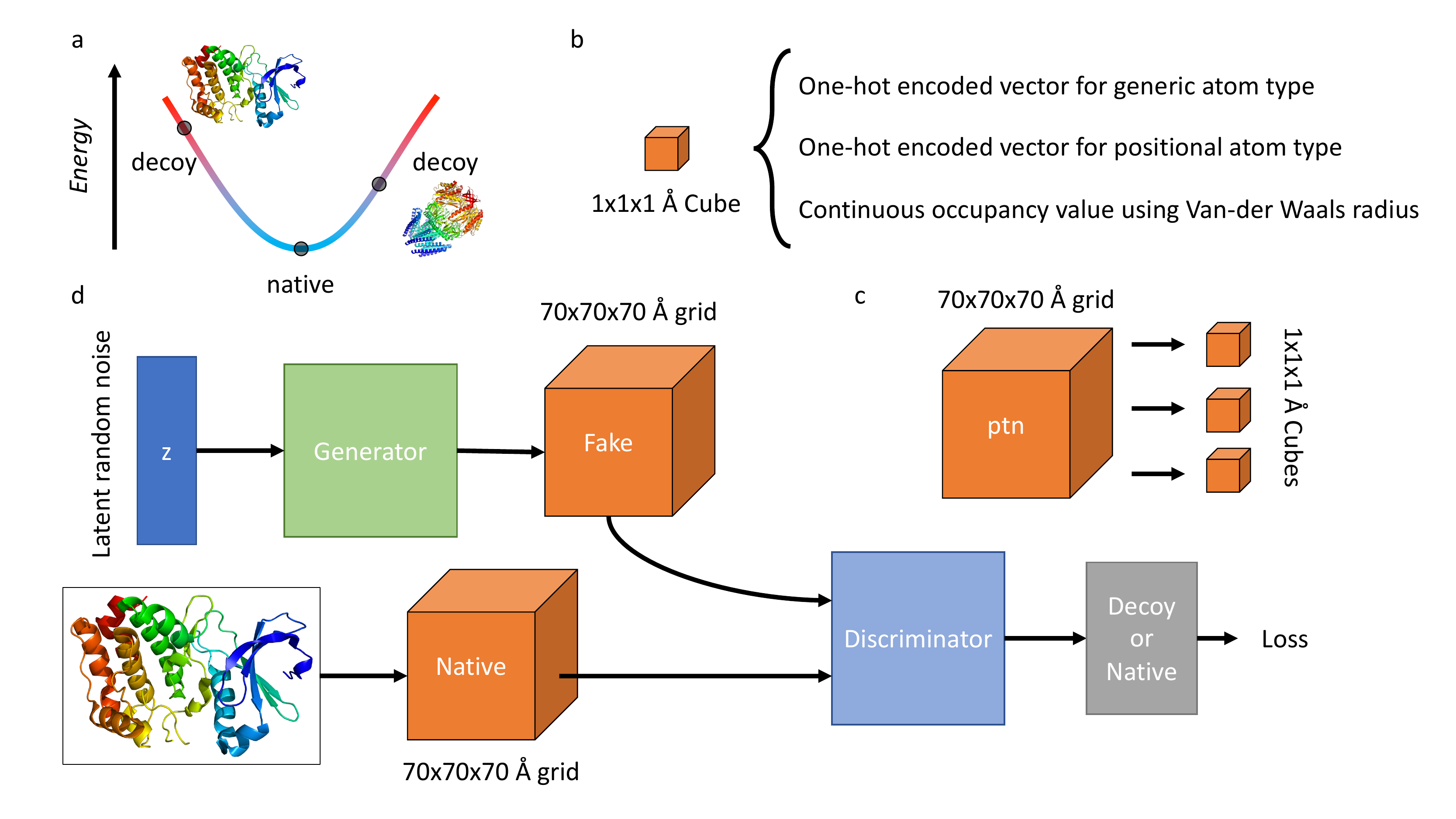}\hfill
    \caption{a) A measurement of deviation of energy of decoy structures relative a respective native structure. b) The features stored in each 1x1x1 angstrom object, representing the space allocated for a single atom. c) The structure of each protein encoded sample. In order to decrease the computational load of the machine learning model, these structures were trimmed down to the smallest non-zero rectangular prism. d) The Generative Adversarial Network architecture in the scope of our work.}
    \label{fig:methods}
\end{figure*}

\subsection{Protein Data Set}

In the Protein Data Bank (PDB) database, every molecular structure can be uniquely identified using a four letter non-case sensitive accession number, also called a PDB ID. 
These molecular accessions are standardized in such a way that the first character is numeric and the last three characters are alphanumeric. An example of such a code is 1crn, identifying a specific hydrophobic protein structure of crambin \cite{teeter1984water}.

Our data set is composed of 1000 proteins segments identified by PDB IDs. Each segment is exactly 11 amino acids long and has more than 80\% alpha helix composition.  

\subsection{Protein Representation}

Protein structures are commonly encoded using a contact map or distance matrix, which is an $nxn$ matrix of pairwise distances between $n$ atoms in a given structure \cite{baldi2003principled}. These structures can be easily used to reconstruct a protein using methods known as multidimensional scaling \cite{kruskal1964multidimensional}. Although this representation captures the distances between each atom, it ignores the atom types which are responsible for forming specific bonds in different levels of protein structure. For instance, in alpha helix secondary structures hydrogen atoms are responsible for holding the helical spiral together. Additionally, sulfur atoms are known to form disulfide brides in the tertiary structure of a protein. Therefore, this work substitutes a contact map for a convolutional network design which has known to be successful in image recognition. Our hope is that convolution can be used in place of contact maps as it uses filter maps to learn the positional relationship amongst features. In addition our use of convolution, in each atom we store three additional features. These features include a one-hot encoded vector for the generic atom type (carbon, nitrogen, sulfur, etc), a one-hot encoded vector for the positional atom type (beta-carbon, alpha-carbon, etc.), and the atomic occupancy value corresponding to the nearest atom given by the following formula: 

$$n(r_{a}) = 1 - e^{-(\frac{r_{vdw}}{r_{a}})^{12} }$$

where $n(r_a)$ is the single atom occupancy of atom $a$ with radius $r_a$, and $r_{vdw}$ is the Van der Waals attractive force radius for atom $a$.

Each protein sample is represented with 70x70x70 Angstrom (Å) grid where each 1x1x1 Å cube contains these three features about a single atom in the structure as displayed in \autoref{fig:methods}b and \autoref{fig:methods}c, respectively.

\subsection{Network Architecture \& Training}

\begin{table}[!htbp]
\label{tab:discriminator}
\centering
 \makebox[0.4\linewidth]{
 \begin{tabular}{|c|c| c|} 
 \hline
 \textbf{Layer type} & \textbf{Output Shape} & \textbf{Param} \# \\ [0.5ex] 
 \hline
 Conv3D  & (19, 13, 16, 64)  & 65728 \\ 
 \hline
LeakyReLU  & (19, 13, 16, 64)    & 0 \\
 \hline
 Conv3D  & (10, 7, 8, 128)  & 221312 \\
 \hline
LeakyReLU  & (10, 7, 8, 128)  & 0  \\
 \hline
Global Max Pooling 3D  & (128) & 0 \\ 
 \hline
 Dense  & (1)    & 129 \\ 
 \hline
\end{tabular}
}
\caption{Model Architecture for discriminator. Discriminator has a total of 287,169 parameters, of which 287,169 is trainable and 0 is non-trainable params}
\end{table}

\begin{table}[!htbp]
\label{tab:generator}
\centering
 \makebox[\linewidth]{
 \begin{tabular}{|c|c| c|} 
 \hline
 \textbf{Layer type} & \textbf{Output Shape} & \textbf{Param} \# \\ [0.5ex] 
 \hline
Dense &  (1169792) &    45621888 \\ 
 \hline
 LeakyReLU   &  (1169792)  & 0 \\
 \hline
 Reshape &  (37, 26, 32, 38)  & 0\\
 \hline
Conv3DTranspose & (37, 26, 32, 38) & 92454  \\
 \hline
 LeakyReLU  & (37, 26, 32, 38)   & 0 \\ 
 \hline
Conv3DTranspose & (37, 26, 32, 38) & 92454 \\
 \hline
 LeakyReLU &  (37, 26, 32, 38)  & 495330 \\ 
 \hline
\end{tabular}
}
\caption{Model Architecture for generator. Generator has a total of 46,302,126 parameters, of which 46,302,126 is trainable and 0 is non-trainable.}
\end{table}

\begin{figure}[!ht]
    \centering
    \includegraphics[width=\linewidth]{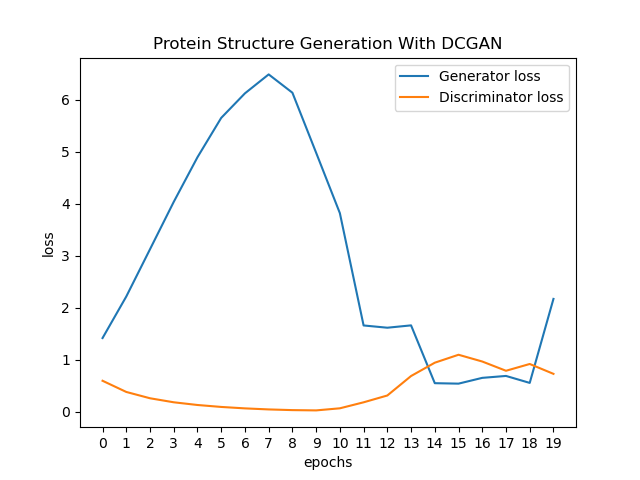}
    \caption{DCGAN training cycle.}
    \label{fig:dcgan_train}
\end{figure}

As it can be seen in figure \autoref{fig:dcgan_train}, we have trained our DCGAN over 50 epochs with an early stopping. The training stopped on epoch around 17 and resulted in a final loss of 2.962 for the generator and 0.642 for the discriminator. Generator has a total of 46,302,126 parameters, all of which are trainable. Discriminator has a total of 287,169 parameters, all of which is trainable. Loss in generator means that the structure prediction is improving over time and the discriminator is trying to discriminate between the decoy and the non-decoy structure generation. Functional annotation generation is robust enough against adversarial attacks.

\section{Conclusion and Future Work} \label{conclusion}

Two optimization steps were included in order to alleviate the computational load of the DCGAN. First, the feature space was trimmed from the 70x70x70 Å window down to the smallest possible rectangular prism grid without removing non-zero occupancy entries. Second, this grid-like formatted data was fed into the DCGAN using a generator function and a batch size of ten.

\section{Experiment Results \& Observations} \label{results}

In this work, we limited our scope to simply discriminating between functional and non-functional protein structures. Although our results provide insight to the difficulty of the problem at hand, a more focused future development would be to create a DCGAN on subsets of protein structure with highly specific functions, such as ligand binding and RNA degradation. Furthermore, adding features such as torsion (Ramachandran) angles between outer bonds of the proteins would increase the representational fidelity of new generated data. Such future work would allow for the possible discovery of novel protein structures that are related to real samples by function. 

\section*{Acknowledgment}

We would like to acknowledge Drexel Society of Artificial Intelligence for its contributions and support for this research. 

\printbibliography
\end{document}